# Dislocation-Solute Cluster Interaction in Al-Mg Binary Alloys

Z. Xu and R.C. Picu
Department of Mechanical, Aerospace and Nuclear Engineering,
Rensselaer Polytechnic Institute, Troy, NY 12180

## Abstract

The close-range interaction of dislocations and solute clusters in the Al-Mg binary system is studied by means of atomistic simulations. We evaluate the binding energy of dislocations to the thermodynamically stable solute atmospheres that form around their cores, at various temperatures and average solid solution concentrations. A measure of the cluster size that renders linear the relationship between the binding energy and the cluster size is identified. It is shown that as the temperature increases the representative cluster dissolves following an Arrhenius law. The corresponding activation energy of this process is found to be independent of the average solute concentration. The variation of the interaction energy between a dislocation and a cluster residing at finite distance from its core is evaluated and it is shown that the interaction is negligible once the separation is larger than approximately 15 Burgers vectors. The data are relevant for the dynamics of dislocation pinning during dynamic strain ageing in solid solution alloys and for static ageing.

**Keywords**: dynamic strain ageing, dislocation mechanics, solid solutions, Al-Mg alloys.

## 1. Introduction

Aluminum alloys are important technological materials. They are used is a large variety of applications including aerospace, automotive and food packaging, primarily due to their advantageous strength to weight ratio, corrosion resistance. The Al-Mg alloys forming the 5xxx series are some of the most widely used. In these alloys Mg is added for strength, its concentration ranging from 2 to 5%. The commercial materials contain in addition to Mg a number of other elements such as Mn, Cu, Si and Fe with concentrations up to 1%. Binary Al-Mg alloys are usually produced for scientific purposes rather than for large-scale consumption.

Al-Mg alloys with Mg concentration on the order or below 5% are solid solutions, with the substitutional Mg solute randomly distributed in the matrix. Although at room temperature a thermodynamic driving force for the formation of the intermetallic phase $Al_3Mg$ exists, the kinetics of its formation is very slow. Nevertheless, Mg is thought to cluster at sources of strain field, such as dislocation cores. The Mg atom is larger than Al by about 12% and hence it produces a relatively short-ranged strain field akin to that of a dilatation center. In presence of another field containing a pressure component, such as that of a dislocation, an interaction develops that leads to a driving force toward or away from the dislocation core.

The clustering of solute at dislocations has been considered for a long time the key component of the mechanism of dynamic strain ageing (DSA). This phenomenon and the associated negative strain rate sensitivity lead to the Portevin-LeChatelier (PLC) effect which manifests itself at the macroscopic scale [e.g. 1,2,3,4]. The PLC effect consists in discontinuous plastic flow. In the temperature and strain rate range in which it is observed, the stress-strain curves exhibit serrations, each serration corresponding to a strain localization event. These leave traces on the surface of the specimen and, upon extensive deformation, lead to surface roughening. More importantly, a significant reduction in ductility (and formability) is observed in the range in which PLC takes place. Obviously, the reduction in formability is detrimental in industrial processes, which provides a motivation for the study of DSA and the associated solute-dislocation interaction process.

The fundamental mechanism of DSA is not fully understood. In broad terms it is believed that solute clustering at dislocations leads to an enhanced resistance to dislocation motion. In the first model proposed by Cottrell and Bilby [5] it was considered that the solute atmospheres travel with the mobile dislocations as long as the defects move at low speed (low strain rate), and is left behind at higher rates. It was later understood that a spatially uniform increase of the friction stress does not lead to serrated flow, and it was proposed [6] that clustering takes place during the relatively short time that mobile dislocations are arrested at obstacles such as forest dislocations, precipitates, etc. Clustering may take place by lattice diffusion, case in which it uniformly locks the dislocation over its entire length, or by pipe diffusion [7], from the decorated obstacle, along the core of the arrested mobile dislocation [8]. If the arrest time is too short for appreciable clustering to take place (high strain rates), the obstacle strength variation is too small to be observed at the scale of the plastic flow (population of dislocations). If the arrest time is very long (in the limit of static ageing), the obstacle strength reaches saturation and is therefore independent of small variations in deformation rate. Hence, negative strain rate sensitivity (higher flow stress at lower deformation rates) results when the two characteristic times, that of the clustering process and the average arrest time of dislocations, are comparable.

Another recently proposed mechanism [9] focuses on forest rather than on mobile dislocations. Similarly with the mobile dislocations, forest dislocations acquire solute clusters. In fact, clustering at forests is more likely as their residence time is much longer than the arrest time of the mobile defects (and this, irrespective of the fact that forests and mobile may exchange roles due to lattice rotations during plastic flow). The effect of clustering at forests is similar to that of clustering at mobile dislocations: it leads to an increase of the strength of dislocation junctions, which are the main obstacles to flow in non-heat treatable alloys. Lower deformation rates lead to longer lifetimes for forests and therefore to larger clusters, which, in turn, lead to higher obstacle strength and flow stress. The opposite situation is expected at higher rates. The result is negative strain rate sensitivity.

Irrespective whether clustering occurs at mobile or forest dislocations, all models of DSA proposed to date [8,9,10,11,12] include a function representing the variation of the obstacle strength with the size of the solute cluster. The strength appears either as a binding energy of the dislocation to the cluster, or as a critical shear stress required for unpinning. Furthermore, as the temperature increases, it is generally considered that either (or both) the cluster dissolves or/and saturation is reached as the diffusion rate is enhanced. In some cases the functions representing these processes are left unspecified and in other cases they are empirical and, to the knowledge of the authors, have not been tested to date against measurement or simulation data.

The objective of the present report is to provide new data regarding the following issues: a) the functional form of the variation with the solute cluster size of the binding energy of a dislocation to the thermodynamically stable solute cluster forming around its core, b) the variation of the cluster size and the binding energy with temperature, c) the effect of the average solute concentration on a) and b), and d) the variation of the interaction energy with the distance between the cluster and the dislocation.

## 2. Modeling and simulation procedure

We consider an Al crystal with 5% of the Al atoms replaced by Mg (5at%). Periodic boundary conditions are imposed in all three directions. In absence of dislocations, the simulation cell is large enough such that the interaction of solute in the cell with their images is negligible. The simulation cell contains a dipole of dislocations, which allows the use of periodic boundary conditions. In addition, the field of the dipole decays faster than that of a single dislocation, which minimizes image effects. Image effects are assumed to be small in the present problem as we are interested only in the structure of the cluster in the immediate vicinity of the core. This is due to the fact that solute atoms in the cluster interact with the core much stronger than they do with their images. After the model is fully relaxed and the two dislocations dissociate, the desired number of Al atoms are replaced by Mg.

A Monte Carlo (MC) procedure is used to evolve the system toward its thermodynamic equilibrium configuration. The moves made are: atom type change, atom displacement by a small fraction of the lattice distance (maximum 0.05 A) modeling lattice vibrations, and volume change of the whole simulation cell (dynamic periodic boundaries). All simulations were performed in the NPT ensemble. The average solute concentration is 1at%, 3at% and 5at%, in separate simulations. The system was equilibrated at various temperatures as specified in the next section.

The model was similar to that used in [13] where the clustering process in the same system, in presence and in absence of dislocations, was investigated. It contains 23958 atoms and has dimensions 88.89 x 174.59 x 25.66 $A^3$, where the 88.89 A dimension is in the direction perpendicular to the dislocation line and contained in the glide plane, while the 25.66 A dimension is measured along the dislocation line. The simulation cell size was selected to be the smallest cell size at which the scaling of the dislocation energy with the cell size begins to follow the log(size) dependence which is expected once short-range effects become negligible. The rates of success of the various MC moves were: the atom displacement move was successful 65% of times, the type interchange was successful 10% of the attempts, while volume changes were made with 20% success rate. The atom type interchange moves have the lowest success rate, as expected. A typical simulation length was of 2 billion MC steps, which corresponds to about 8000 successful interchange moves per atom.

The total energy of the system as well as the size of clusters was monitored during the run. Data were collected only after the system was fully equilibrated and those parameters monitored reached a stable running time average.

Once the stable cluster at the given temperature was obtained, molecular statics (MS) was used to determine the binding energy of the dislocation to the cluster. The procedure requires determining the energy of the four systems shown in figure 1. The binding energy results as

$$E_b = (E_{i1} - E_{f2}) + (E_{i2} - E_{f1}), \qquad (1)$$

where *i* and *f* stand for “initial” and “final” states, respectively.

Let us consider an initial configuration with the dislocation being clustered (initial configuration i1), figure 1a. After unbinding the dislocation from its solute cloud, the defect moves “at infinity,” into a region of the material with no cluster and which, before unbinding, contains just a random distribution of solute (initial configuration i2), figure 1b. The cluster is left behind. The configuration corresponding to the initial region i1 in figure 1a becomes the final configuration f1 shown in figure 1c. That corresponding to the initial region i2 in figure 1b, becomes the final configuration f2, figure 1d. In order to find the binding energy, one has to evaluate the difference between the final and the initial configurations.

The difference in energies of i1 and f2 is the energy variation during the MC clustering simulation between the initial and final configurations. A large number of replicas were generated for configuration f2, and $E_{f2}$ was evaluated as the average energy of those replicas. MC simulations starting from f2 yield configuration i1. The simulation also provides the Mg concentration field, i.e. the solute concentration at every site.

The energy of f1 can be obtained by performing relaxation of the system i1 after the dislocation has been removed. To compute $E_{f1}$ we generate a large number of replicas by simply sampling the concentration field obtained from the MC clustering simulation. The replicas contain no dislocation and a thermodynamically unstable cluster. The cluster needs to be slightly shifted as the dislocation is removed from the model. We note that this is not an unphysical alteration of the cluster. The transformation is identical to what happens in the crystal when the dislocation moves out under the action of a stress field, i.e. a relative shift of atomic planes across the glide plane by one Burgers vector. The energy of the system i2 is straightforward to evaluate as this configuration contains just a random distribution of Mg.

This procedure leads to the exact cancellation of all energetic interactions between the two dislocations in the model, the dislocations in the model and their images, between the two clusters and the clusters and their images, which should not enter the binding energy.

This procedure requires extensive averaging over a large number of configurations in order to eliminate the noise associated with the random distribution of Mg atoms and to extract the statistically relevant quantities. Replicas were run on a parallel machine using up to 8 CPUs. Two thousand configurations are used for each figure reported.

The MC and MS simulations were run by using an embedded atom potential developed for this binary system by Liu *et al*. [14]. The most used potentials for pure Al are those developed by Ercolessi and Adams [15] and by Mishin *et al* [16]. An EAM potential for pure Mg was fitted by Liu *et al* [17]. The potential used here was developed based on the Ercolessi-Adams functions and fit to a number of physical data, such as the dilute heat of mixing of Mg in Al and of Al in Mg, the activation energy for diffusion of Mg in Al, the lattice constant change of Al upon addition of Mg, and the heat of formation of several Al-Mg intermetallic phases. The dilute heat of mixing and the heat of formation of the various phases were obtained from LDA/DFT pseudopotential calculations. Additionally, it was fit to a large database containing interatomic forces evaluated by *ab initio* methods. The potential was previously used to study grain boundary and surface segregation in Al-Mg solid solutions [14,18], Mg clustering in the dislocated and un-dislocated Al crystal [13] and dislocation mobility in Al-Mg alloys [19].

## 3. Results and discussion

We first derive a measure of the cluster size that is convenient to use in the present context. Then we determine the binding energy as a function of cluster size and the effects of temperature and average solute concentration on the binding energy. Finally, we investigate the variation of the interaction energy with the distance of the dislocation to a cluster of given size and we calibrate an equation to describe this type of interaction.

The discussion here focuses on a $60^o$ dislocation on which no external stress is applied. The selection of this type of dislocation was dictated by our previous work on the mechanism of DSA. In the model developed in [9], a Lomer-Cottrell lock in which both the forest and the mobile dislocation are of $60^o$ type is considered. The forest dislocation carries a solute cluster. The strength of the lock was determined for various magnitudes of the binding energy of the forest to the cluster. This function enters a mesoscopic model which is used to predict the negative strain rate sensitivity of the flow stress. The dependence of the binding energy on the cluster size is an unspecified function in Ref. [9]. In a later development of that model, a large number of junction types were considered and the analysis was repeated for each. It was observed that the behavior of the mesoscopic model for the flow stress based exclusively on the symmetric Lomer-Cottrell lock [9] was representative for the whole population of junctions [20]. Based on this observation, we derive here the dependence of the binding energy (Q in [9]) on the cluster size and temperature, the function missing from the analysis in [9,20].

*3.1 Measure of the cluster size*

The structure of the thermodynamically stable cluster around the core of edge, screw and $60^o$ dislocations in Al-Mg was discussed in [13]. Here we are only interested in deriving a measure of the cluster size that may be used as a scalar variable in the function of the binding energy.

Figure 2 shows the histogram of Mg concentration in the model of the $60^o$ dislocation at room temperature and at $473^oK$. The curves were obtained by averaging over 8 replicas. The curve corresponding to the model without the dislocation (unclustered; curve not shown) is a rather narrow Gaussian centered at the average solute concentration (5%). Once clustering takes place, the peak at 5% splits in 2 peaks, one at lower concentration, corresponding to the region above the glide plane which is depleted of solute, and one at higher concentration. The second peak indicates the tendency to form the $Al_3Mg$ phase in the region of higher Mg concentration below the glide plane. The structure of the phase is discussed in [13]. Mg atoms are located at every second site along <110> directions in {111} planes, while every second {111} plane in the sequence contains Al only. The structure corresponds to 12.5at% Mg, which is the concentration at which the second peak is observed in the curve corresponding to room temperature. At the elevated temperature, the high concentration peak moves closer to 5% (at 6.5at%) indicating disordering of the intermetallic phase.

The measure of the cluster size used here, $s$, is the first moment of the probability distribution function (PDF) in figure 2 computed only from the region with solute concentration larger than the average $<c>$ ($<c>$ = 5% in figure 2), specifically:

$$s = \int_{<c>}^{\infty} PDF(c)(c - <c>)dc = \frac{1}{N}\sum_{i=1}^{N} H(c_i - <c>) \qquad (2)$$

where $H$ is step function at zero, $c_i$ is the Mg concentration at site $i$, and N is the total number of atoms in the simulation cell. A random distribution of solute leads to $s$ = 0 (after averaging over replicas).

*3.2 Dislocation-cluster binding energy, effect of temperature and average Mg concentration*

Clusters having various sizes were prepared using the MC procedure described in Section 2. All of them are stable, each corresponding to a different temperature. Apparently, clusters of various sizes could have been obtained by interrupting the MC equilibration. We chose not to use this procedure since the cluster growth process in the MC simulations does not follow the real dynamics and hence the resulting clusters would have no relation with real ones.

The variation of the binding energy per unit length of dislocation line, $E_b$, with the cluster size is shown in figure 3. The variation is linear in the range of small clusters and departs from linearity at very large clusters. The largest cluster considered ($s$ = 0.015) is stable at room temperature and is similar to that presented in [13].

This linear variation allows incorporating kinetics in the prediction of the binding energy. The kinetics of solute clustering may be described using the original Cottrell-Bilby model [5] modified by Louat's correction [21] to account for saturation. The time variation of the concentration at the core is described by:

$$s(t) = s_\infty\left(1 - \exp\left(-\left(t/\tau\right)^n\right)\right) \quad (3)$$

where $s_\infty$ is the cluster size at saturation, and $\tau$ is a time constant associated with diffusion. $\tau$ depends on the diffusion coefficient in the vicinity of the core and on the local solute concentration. The exponent $n$ is usually taken 2/3 if bulk diffusion dominates and 1/3 if clustering takes place by pipe diffusion.

Clustering in the vicinity of the core is complicated due to the fact that close to the core diffusion takes place in a region with large strain and concentration gradients. Hence the effective diffusion coefficient becomes position dependent. The kinetics of clustering may be studied by kinetic MC or by using a continuum model in which the constants are calibrated from atomistic study. Such analysis would validate equation (3). This work is in progress and will be presented in a separate report.

Using equation (3) and the linearity of the curve in figure 3, one may write

$$E_b(t) = E_{b\infty}\left(1 - \exp\left(-\left(t/\tau\right)^n\right)\right) \quad (4)$$

which is valid for sufficiently small cluster sizes ($s$ <0.008). Note that $s$ = 0.008 corresponds approximately to the cluster which is thermodynamically stable at 373$^{o}$K.

Figure 3 combines data for three average solute concentrations, 1at%, 3at% and 5at%. Only three cluster sizes, corresponding to temperatures 373K, 473K and 673K were considered for the two lower average concentrations. When using the measure of cluster size in equation 2, all data follow approximately the same dependence on the cluster size, $s$. It is noted that the cluster size at given temperature decreases with $<c>$, as expected.

The variation of the binding energy with temperature is shown in figure 4. The plot has a linear region that corresponds to temperatures larger than 373$^{o}$K for $<c>$ = 5%. This coincides with the region in which $E_b$ is proportional to the cluster size in figure 3. The variation is linear for all temperatures investigated at lower average concentration. In the linear region, the dependence may be described by an Arrhenius law. The slope of this linear region is ~ 0.0033 eV per A of dislocation line, identical for all three concentrations. This quantity is representative for the growth/dissolution of the cluster at given temperature and should not be interpreted as an activation energy for the separation of the dislocation from the cluster. The independence of the slope from the average solute concentration is expected, as the activation energy for cluster dissolution should be a function of the dislocation type/Burgers vector only.

Several comments regarding the relevance of these results for the mechanism of DSA are in place. In DSA, whether clustering takes place by bulk or pipe diffusion, at mobile or forest dislocations, it is rarely expected that clusters reach saturation. Negative strain rate sensitivity is observed in Al-Mg at temperatures ranging approximately from 180$^{o}$K to 370$^{o}$K [22,23] and at strain rates below $10^{-1}$ $s^{-1}$. The strain rate sensitivity parameter reaches its minimum at about 323$^{o}$K. At this temperature, Mg diffusion is rather slow and largely insufficient to produce cluster saturation during the mobile dislocation arrest time (on the order of 0.1 s, depending on the strain rate) or the residence time of forest dislocations (about 50 times longer than the arrest time of mobile dislocations). The present data show that since clusters are expected to be rather small, a time and temperature dependence of the type

$$E_b(t,T) = E_{b0} \exp\left(-\frac{E_a}{kT}\right)\left(1-\exp\left(-\left(t/\tau(T)\right)^n\right)\right) \quad (5)$$

may be used in DSA models. The average solute concentration <$c$> enters only through the pre-exponential factor $E_{b0}$. The kinetics of clustering represented by the characteristic time τ(T) is also expected to depend on <$c$> to some extent.

Another insight refers to the question whether cluster dissolution may lead to the increase of the negative strain rate sensitivity parameter to positive values as the temperature increases above 373°K (cessation of the PLC effect). The answer appears to be negative. No dramatic change is seen in figure 4 in the vicinity of this temperature; the stable cluster size variation with temperature is rather gradual.

### *3.3 Long-range dislocation-cluster interaction*

It is further interesting to investigate the situation in which the dislocation is located at some distance away from the cluster. Let us consider the configuration in Fig. 5. The situation discussed in the previous Section is denoted here by A: the dislocation field has driven the formation of the cluster and a region of enhanced solute concentration develops below the glide plane. We are now interested in configuration B and the variation of $E_b$ with the distance $y$. Note that B is mechanically stable as the dislocation is assumed not to climb and the energetic driving force for glide vanishes.

The binding energy is computed by using the procedure presented in Section 2 and the analysis is repeated for various $y$ and various cluster sizes. The cluster is obtained by first placing the dislocation in configuration A, running the MC simulation for clustering, removing the dislocation from A, re-inserting it at B and relaxing.

Figure 6 shows the variation of $E_b$ with $y$ (normalized here by the Burgers vector, $b$ = 2.851A). The variation may be described by a relation of the form:

$$E_b(y) = m + \frac{n(T)}{y+8}, \quad (6)$$

where $m$ and $n$ are fitting coefficients.

Let us discuss the expectations for equation (6). The interaction of the dislocation with the cluster is mechanical in nature if the core is outside of the cluster. It is due to the fact that the cluster has an intrinsic dilatation as well as to the change in the local elastic constants of the lattice in the cluster region.

The interaction energy between a dislocation and a dilatation center depends on the distance $y$ separating the two singularities as $1/y$, i.e. is proportional to the magnitude of the dilatation center and the value of the pressure produced by the dislocation at the location of the center of dilatation [24]. The interaction energy of the dislocation with an inhomogeneity has a similar variation with $y$ in the leading term [25]. Hence, equation (6) would have been expected to be of $1/y$ form.

It is useful to consider now the situation when $y$ is comparable with the dimensions of the cluster. It has been shown [26] that the interaction energy between a sphere of radius $a$ containing a distribution of sources, and a point source of field located at distance $y$ from the surface of the sphere, both located in an infinite body, may be written as the interaction energy between two point sources of field separated by distance ($y+a$) multiplied by a coefficient which depends on $a$. Hence, the proper relation for the interaction energy in the problem discussed here must be of the form (6). The constant $a$ = 8 was obtained by fitting to the computed $E_b$ vs. $y$. It was checked that indeed, $a$ = 8 corresponds to a point approximately in the middle of the simulation cell, which is also approximately the center of the effective cluster. If $y >> a$, equation (6) becomes simply $1/y$, as expected for the interaction between point sources. However, as can be seen in figure 6, the dislocation-cluster interaction becomes negligible when $y > 15$ and therefore only the present form appears to be relevant (equation 6).

The coefficient $n$(T) in equation (6) is approximately proportional to the total dilatation of the cluster and, as this quantity depends on the cluster size, is temperature dependent. Hence one may write

$$n(T) \cong 520s(T) = 520s_0 \, exp\left(-\frac{E_a}{kT}\right). \qquad (7)$$

where the Arrhenius term is identical to that in equation (5) and represents the dependence of the cluster size on temperature.

## 4. Conclusions

The binding energy of dislocations to solute clusters forming around their cores during static or dynamic ageing was studied in Al-Mg alloys using a chemistry-specific atomistic model. The analysis provides the functional form of the binding energy-cluster size relationship, expression required in all current models of static and dynamic strain ageing. The analysis of the longer-range interaction of the dislocation with a cluster shows that the interaction effectively vanishes when the distance between the two entities is larger than 15 Burgers vectors. At shorter distances, the interaction energy scales as the inverse of the distance, provided it is measured from the geometric center of the cluster. This information is valuable in studies of the interaction of mobile dislocations with fluctuating clusters during plastic flow. This type of analysis is also relevant for dynamic strain ageing in these alloys.

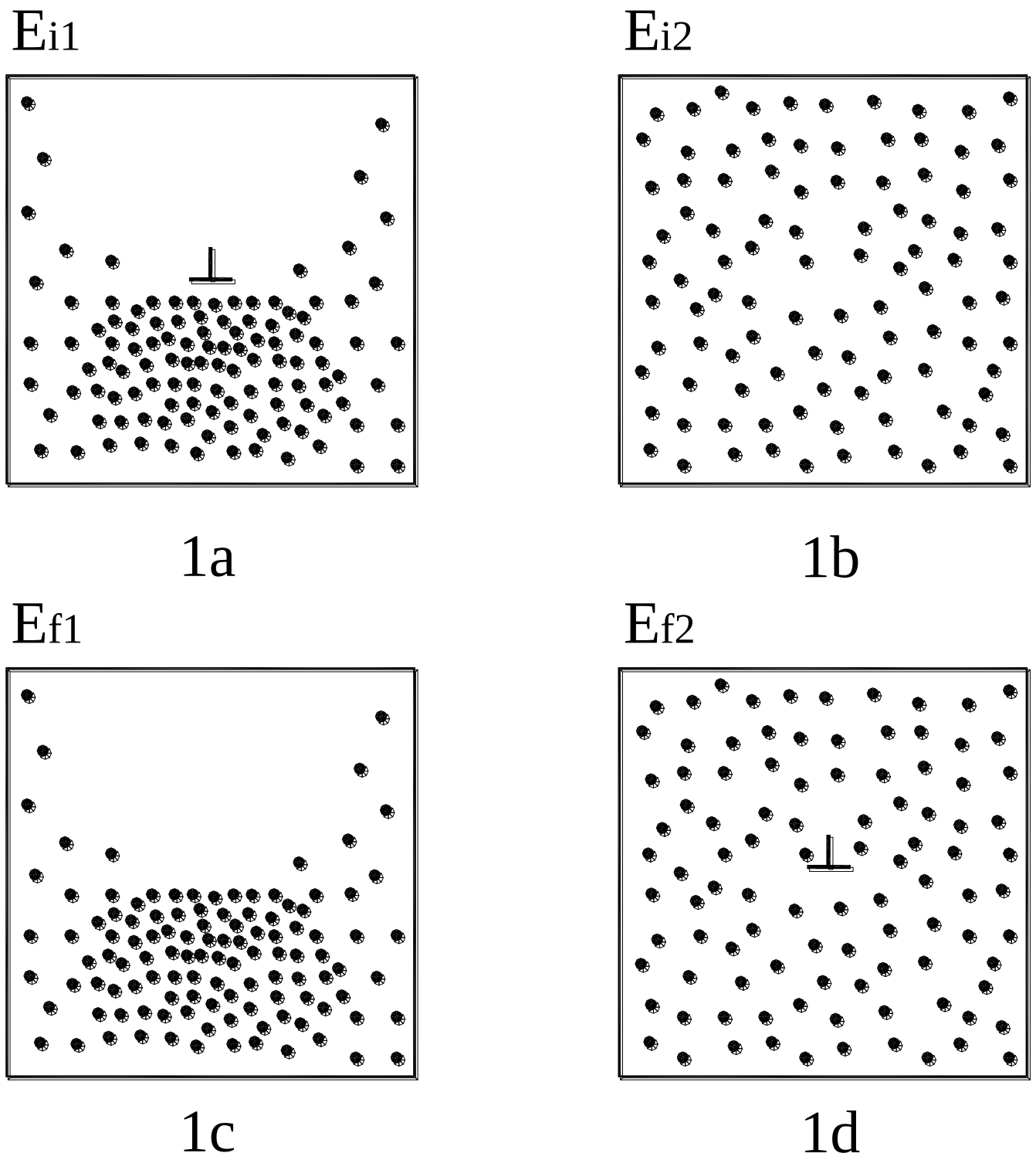

Figure 1. Four configurations used to evaluate the biding energy between a dislocation and the thermodynamically stable solute cluster forming around its core.

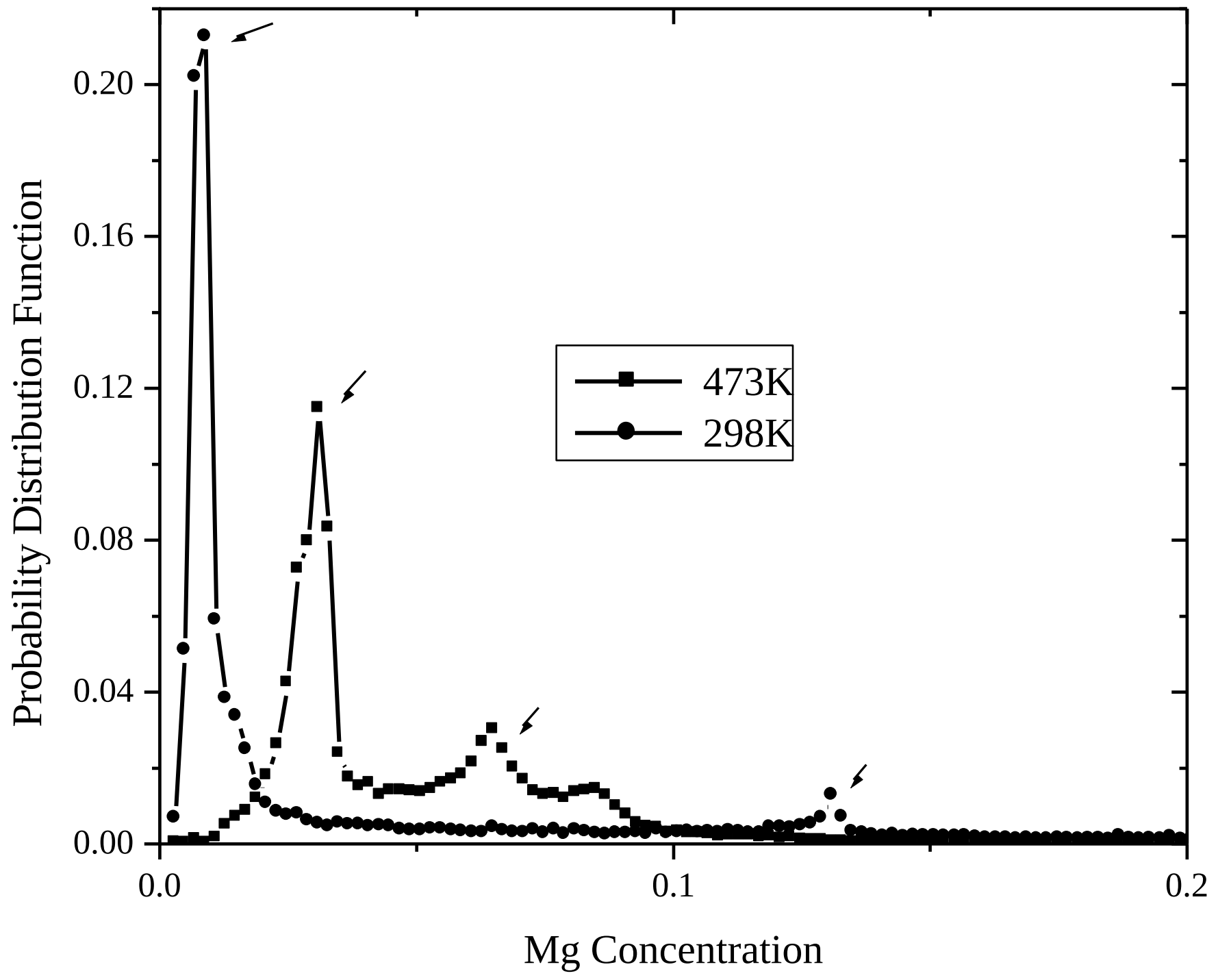

Figure 2. Distribution function of Mg concentration at room temperature and 473°K in a system with average <c> = 5at% Mg. The arrows point to the peaks of the two distributions, one below and one above <c>. The peak at large concentrations and 298°K corresponds to the phase $Al_3Mg$ that forms close to the core.

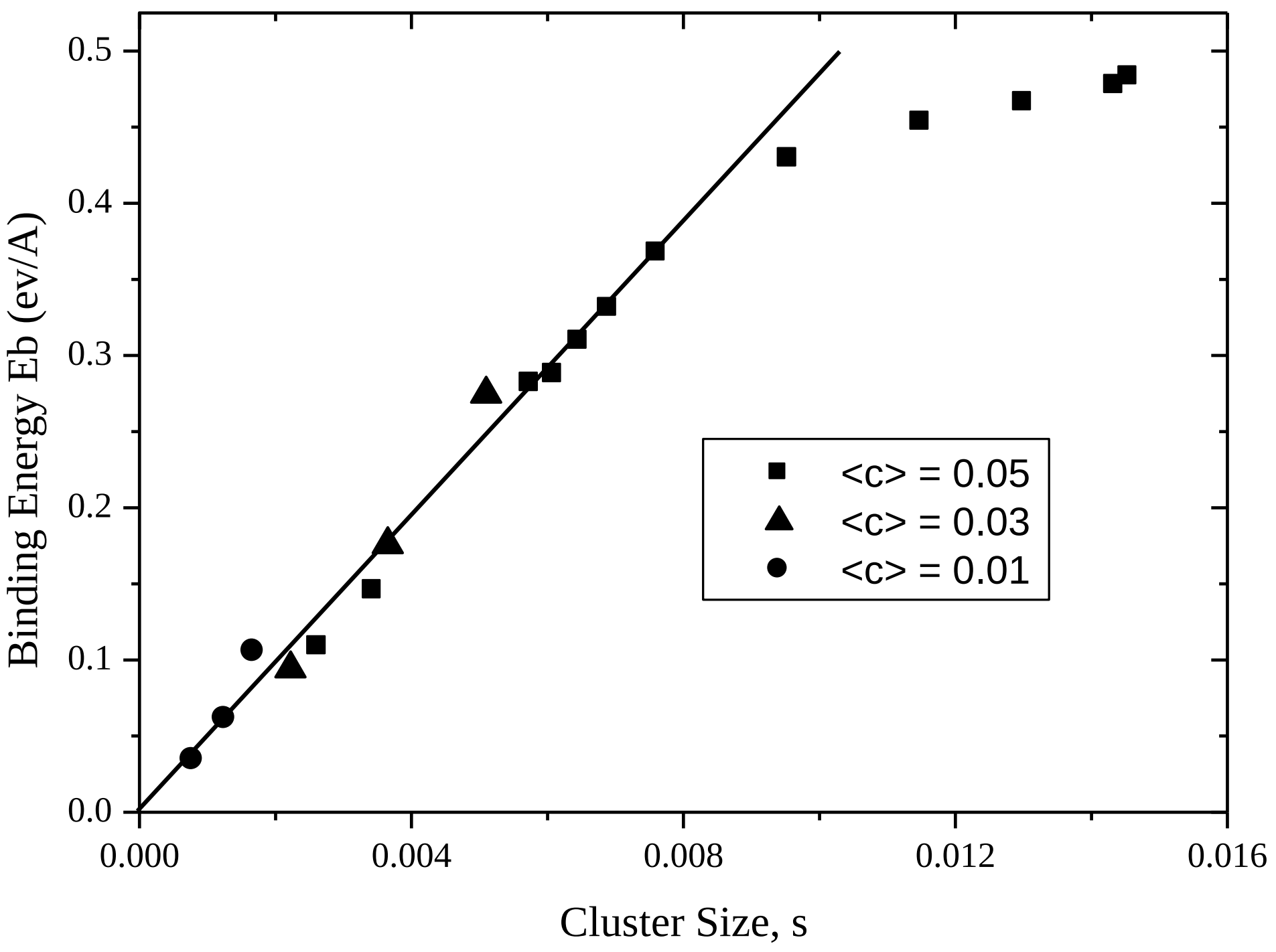

Figure 3. Binding energy per unit length of dislocation line as a function of the cluster size for three values of the average solute concentration. Only three cluster sizes, corresponding to temperatures 373°K, 473°K and 673°K were considered for the two lower average concentrations (circles and triangles).

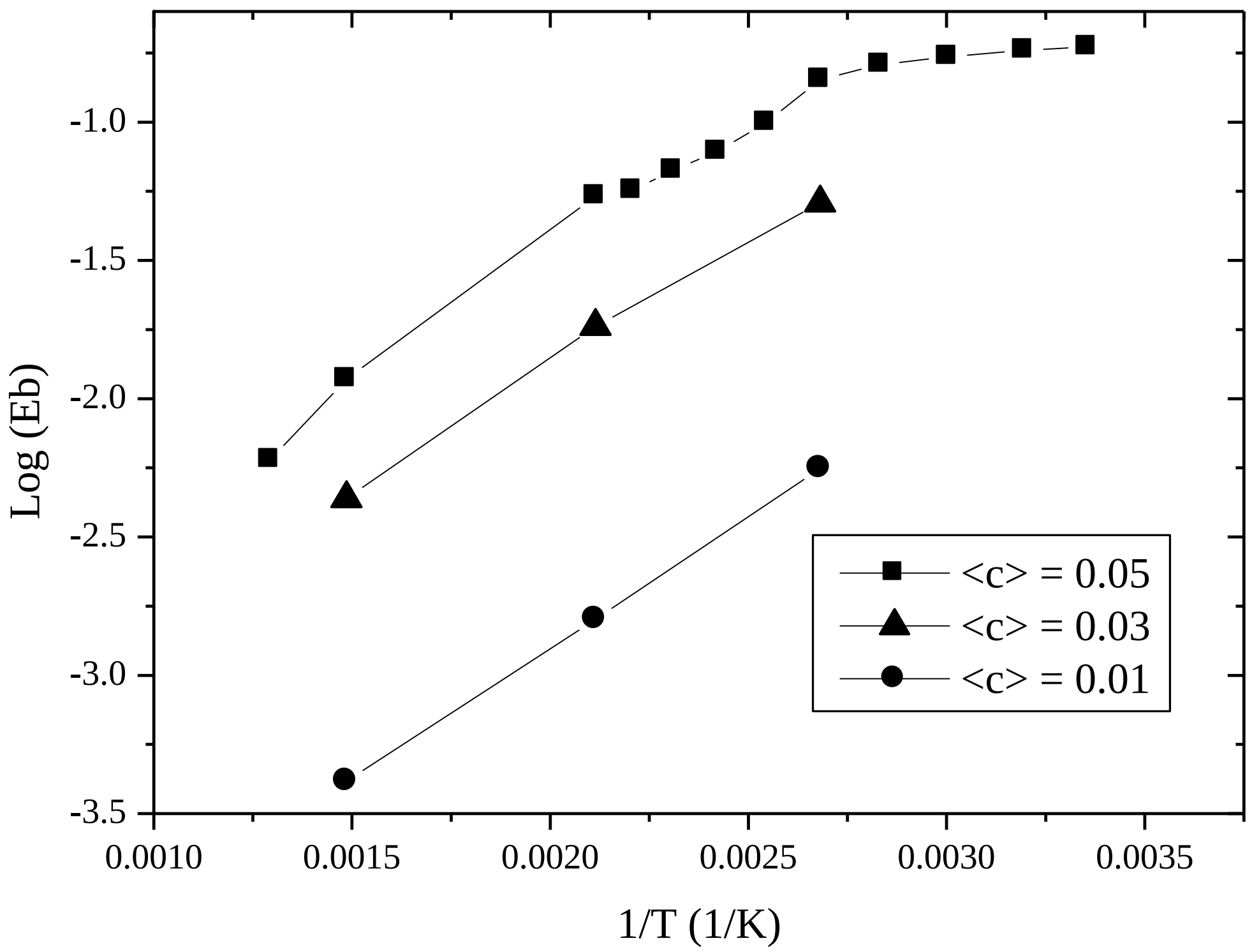

Figure 4. Binding energy of a 60° dislocation to its thermodynamically stable cluster for various temperatures and average solute concentrations.

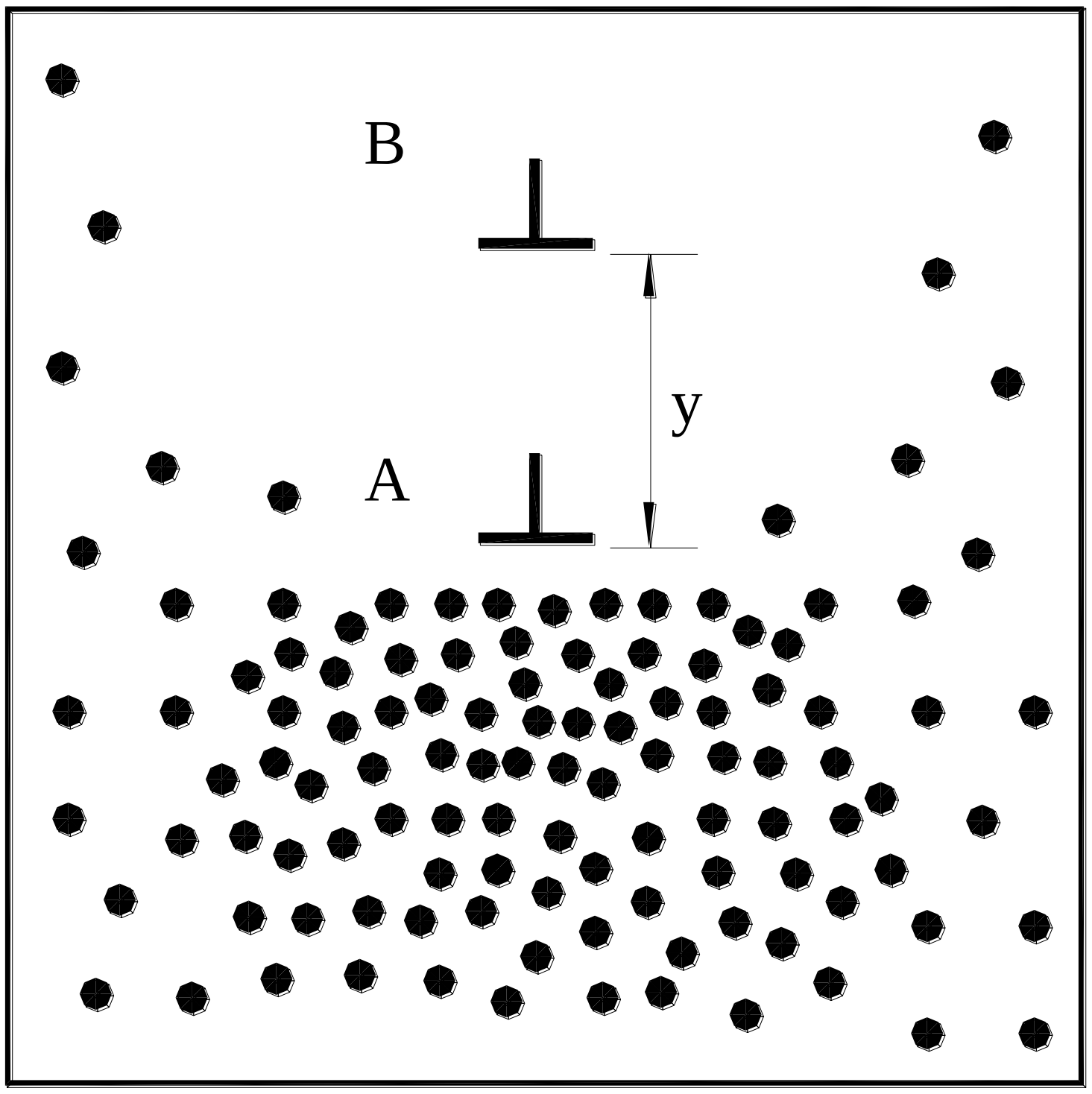

Figure 5. Sketch of configurations used to evaluate the dependence of the binding energy on the dislocation-cluster distance, *y*.

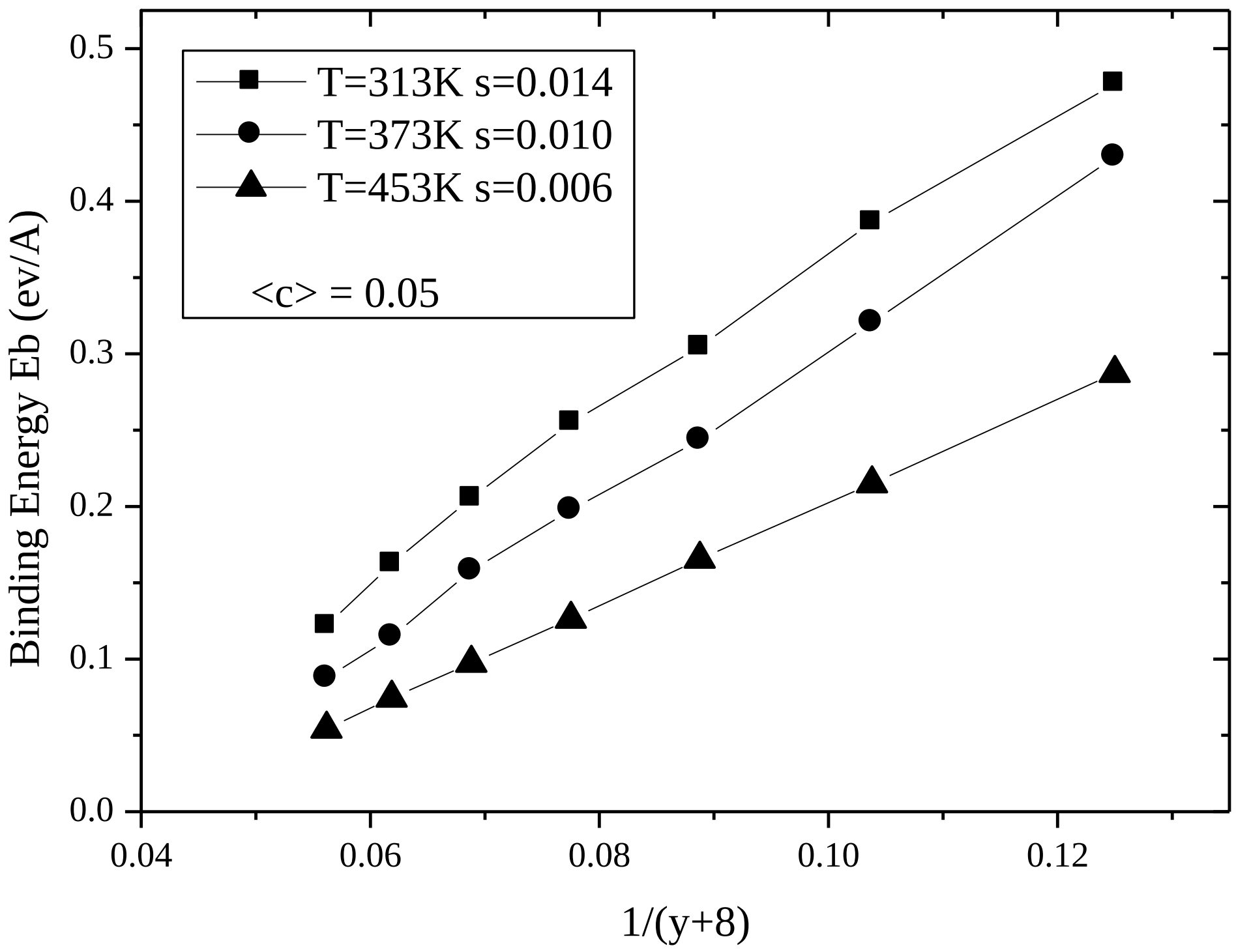

Figure 6. Binding energy vs. the dislocation-cluster distance for three cluster sizes, at average solute concentration <*c*> = 5%. The distance *y* is normalized by the Burgers vector length.